# The role of natural language in understanding the universe: a teaching-learning sequence for high school students

M. Tuveri[1,2], V. Fanti[1,2]

[1] Physics Department, University of Cagliari, Cittadella Universitaria di Monserrato, 09042 Monserrato (CA), Italy

[2] Istituto Nazionale di Fisica Nucleare, Sezione di Cagliari, Cittadella Universitaria di Monserrato, 09042, Monserrato (CA), Italy

**Abstract**

Introducing gravitational physics at high school provides educational means for bridging the gap between the image of science held by students and science itself. Natural language is fundamental in this learning. It engages students in constructing an understanding of a concept or a notion, establishing new relations between previous and new elements of knowledge. We present a teaching-learning sequence (TLS) aimed to contextualize gravitational physics along the lines of the Einstein Telescope educational program in Sardinia devoted to upper secondary school students. We focused on the role of debates and controversy in the evolution of science, proposing science-reflexive meta-discourses to present physics as a unified knowledge textbook. We discuss our design and present results by analyzing students' semiotic registers to recap their learning during the activity. Finally, we discuss the potentiality of our TLS in orientating students towards STEM.

**Introduction**

Since the dawn of human inquiry into the cosmos, the study of the universe has been driven by a desire to understand the fundamental forces that shape it, particularly gravity. The development of modern cosmological theories and the rise of gravitational physics have profoundly transformed our understanding of the universe. Central to this shift was the birth of Einstein's theory of general relativity and the emergence of modern cosmology, which revolutionized our conception of space, time, and gravitational forces [1-3]. Throughout history, cultures have sought to explain the natural world through cosmological models. However, it was only with the advent of general relativity and the discovery of gravitational waves that a deeper understanding of the universe's structure and dynamics began to take shape [4-9]. These groundbreaking theories, which describe how massive objects influence spacetime, have significantly expanded our view of the cosmos [4]. Gravitational waves, in particular, have opened a new window into the universe, providing a direct means of observing cosmic events like black hole mergers, thus pushing the boundaries of both theoretical and observational physics [10]. Future detectors, such as the Einstein Telescope (ET) will be capable of observing the entire Universe using gravitational waves aiming to increase a factor of ten the sensitivity of previous generation detectors [11-13]. The historical development of these theories offers invaluable insight into the scientific process, making it an ideal focus for educational activities aimed at high school students [2,3]. Exploring the historical development of these theories offers students an opportunity to engage with the very origins of modern physics. It enables them to grasp not only the scientific concepts behind general relativity and cosmology but also the process through which scientific ideas evolve [14,15].

However, misconceptions—such as the widespread misunderstanding of the "Big Bang" as an explosion—can hinder students' comprehension. These misconceptions, often stemming from intuitive but incorrect visualizations, obscure the actual physics involved, particularly when considering the expansion of spacetime itself. In contemporary physics, these misunderstandings largely result from incorrect interpretations of metaphorical meanings [16,17]. Addressing these misconceptions is a priority for educators, as it encourages students to critically assess the nature of scientific theories and their historical development [14,15]. Promoting conceptual learning through a proper understanding of metaphors is essential, as metaphors play a crucial role in learning, influencing how problems are conceptualized and how solutions are approached [18-20]. They are not just linguistic tools but shape the types of actions and strategies deemed appropriate for addressing challenges. Constructing an understanding requires connecting elements of knowledge, with active engagement and mental involvement being key. Semiotics and language, particularly natural language, are critical in learning, especially when transitioning between different registers [21]. For istance, Duval emphasized the importance of relating these registers to construct meaning [22]. This is particularly true in in complex domains like quantum mechanics [23].

Research in physics education (PER) underscores the importance of presenting a unified perspective on the evolution of physics and the key figures who contributed to the development of scientific knowledge [24,25]. Such an approach highlights that "scientific revolutions" and the notion of "genius ideas emerging from nowhere" are better understood as metaphors rather than historical truths. As Heilbron (see [15], p. 8) emphasized, "The Scientific Revolution is a metaphor as well as a shorthand for particular developments within a historical period. The metaphor has given rise to much debate between those who regard these developments primarily as an abrupt transformation and those who view them as a continuity of pace and content." This viewpoint clarifies that the "Scientific Revolution" is both a metaphor and a shorthand for specific developments within a historical context. The debate surrounding this metaphor reveals two perspectives: some see these developments as abrupt transformations, while others emphasize continuity in both pace and content.

Placing scientific research within its historical context is an effective way to highlight the controversies, challenges, questions, and answers that have shaped scientific progress [26]. By exploring the historical evolution of gravitational physics, instructors can stimulate curiosity and motivate students to engage deeply with the topics at hand. This approach has been shown to combat misconceptions in science learning and foster conceptual understanding [3]. Indeed, numerous studies have demonstrated the benefits of contextualizing physics education, highlighting its positive impact on students' learning experiences. This method not only strengthens students' grasp of scientific concepts but also enhances their attitudes toward the nature of science. It promotes skills in scientific argumentation, metacognition, and a deeper understanding of core scientific ideas [17,24-26].

The aim of this study was to design a teaching/learning sequence focused on the historical and epistemological aspects of contemporary physics, particularly general relativity, quantum mechanics, and modern cosmology. The goal was to help students better understand these complex concepts, addressing the following research question:

RQ1: How can we design a teaching/learning sequence that integrates the historical and epistemological reconstruction of cosmology (relativistic astrophysics) with modern research in physics in a non-formal learning context?

RQ2: Is it possible to monitor the learning process of students during such a learning activity?

RQ3: What mental associations and conceptual categories do students develop during the process of learning physics in a non-formal context?

**Theoretical framework**

In recent years, there has been a growing global interest in diversifying scientific curricula in secondary school to enhance students' learning experiences, particularly in physics. This shift allows a broader range of topics to be explored, directly addressing learners' interests. In particular, non-formal education appears to hold great potential in this regard, as it provides opportunities for educational experimentation aimed at promoting the learning of physics in schools, motivating, intriguing, and capturing the attention of students, teachers, and the wider public [26]. Non-formal education occurs outside traditional schooling but within an institutional framework, such as cultural centers or extracurricular activities, and is often delivered by specialists. It fosters interdisciplinary skills, flexibility, and active learning, providing opportunities for personalized learning strategies based on students' interests. This approach complements formal education by promoting holistic learning, offering avenues for self-assessment, and increasing student engagement [27,28].

Teaching/learning sequences (TLS) are particularly well-suited for this context [29-36]. They consist of structured instructional activities designed to guide students through the exploration and understanding of specific scientific concepts. These sequences incorporate progressive learning steps that include conceptual development, student engagement, and formative assessment, all aimed at promoting deep understanding and meaningful learning. TLS ensures the sequentiality and coherence of lessons, emphasizes key concepts, and incorporates active, playful, and collaborative methodologies.

The TLS developed for this study was grounded in the principles of Design-Based Research (DBR) and the Model of Educational Reconstruction (MER) [37-42]. These methodologies provide a comprehensive framework for constructing, testing, and refining both educational content and strategies. They enabled the continuous development of scientifically accurate material alongside pedagogically effective practices, ensuring the sequence's suitability for high school students studying complex topics such as cosmology and relativistic astrophysics. Developing teaching strategies and methodologies that can be directly applied in the classroom is crucial for fostering effective learning environments. Experimenting with practices that actively engage students, including collaborative, inquiry-based, and play-based learning approaches, has been shown to be particularly promising in informal and non-formal contexts. In particular, Inquiry-Based Science Education (IBSE) offers an effective pedagogical framework, allowing students to participate actively in their learning process through hands-on and minds-on activities.

Table 1. Sample, topics, and structure of our TLS

| Sample | N | Duration | Topics and concepts | Learning tools |
| --- | --- | --- | --- | --- |

| Total | 119 | 90 minutes | Motion of bodies | Experiments on the fall of heavy bodies (e.g., common objects, balls) to work on the concept of force and related conceptions |
| --- | --- | --- | --- | --- |
| Males | 68 | | | |
| Females | 51 | | • Aristotelic motion (natural motion, horror vacui) | |
| Classes | | | • Newtonian dynamics | |
| 3rd | 45 | |     o Force, acceleration, Equilibrium (F=ma) | Educational videos and animations on free fall, motion on curved spaces |
| 4th | 58 | |     o Gravitational force | |
| 5th | 16 | | • Einstein's gravity | |
| Groups | 36 | |     o Equivalence principle | Debates, questions, and answers. |
| | | |     o General covariance | |
| | | |     o Spacetime and matter | |
| Schools | 36 | 60 minutes | The universe | «embodied» experience on gravitational lensing |
| Artistic | 1 | | • Aristotle's cosmology (geocentrism) | |
| | | | • Pre-relativistic universe (heliocentrism) | Debates, questions, and answers |
| Applied Sciences | 12 | | • Relativistic astrophysics | |
| | | |     o Cosmological model | |
| Scientific | 106 | |     o Big Bang metaphor | |
| | | |     o The observable universe | |
| | | |     o The dark universe | |
| | | |     o GR tests | |
| | | |     o Gravitational waves | |
| | | |     o Interferometers and the Einstein Telescope | |
| | | 30 minutes | Formative evaluation | Wording, and crossroads |

Table 2. Structure of the formative evaluation tasks.

| Task | Description | Teaching/Learning benefits |
| --- | --- | --- |
| **Wording ("Today's words")** | Final summary activity to keep track of students' reconstruction of all the contents discussed during the meeting. | • Promoting instructor-student and student-student interactions<br>• Promoting creativity: students explain why they chose a certain word; in what context it was used and why it summarizes a certain content covered<br>• Construction of conceptual maps – creation of nodes and mental associations between learning contents and real-world phenomena |
| **Crossroads ("Revolutionary crossroads")** | The students, divided into groups, choose one or more words and give a definition. The words and definitions are collected by the teacher who presents the individual definitions to the class through the crossword game. The group that guesses the most words (excluding those they wrote) wins. | • Promoting cooperative learning and inquiry through a minds-on activity<br>• Possibility for the instructor to uncover any erroneous conceptions and investigate whether they are real or dictated by the extreme synthesis imposed by the game.<br>• Promoting the ability to create a link between physical knowledge and natural language. |

**Methodology**

The project, which targeted four schools in Sardinia with students from the third, fourth, and fifth years (from 16 to 19 years old, respectively), was part of the Italian PNRR initiative "Le frontiere della fisica" (The Frontiers of Physics), organized by the Physics Department at the University of Cagliari. The activity aimed to introduce students to gravitational physics, specifically focusing on gravitational waves and Einstein Telescope physics. It consisted of four meetings, each lasting three hours, and scheduled between January and February 2024. Each session was structured with 150 minutes of content delivery. Students worked in small groups to encourage collaborative learning and engagement with the material. A description of the sample can be found in Table 1.

The design of the TLS began with a thorough analysis of the scientific content related to these topics, which was then incorporated into the sequence within a cooperative learning framework. This approach allowed for iterative refinement, ensuring the content remained both scientifically accurate and accessible to students. Emphasis was placed not only on the scientific accuracy of the concepts but also on presenting them in ways that would enhance understanding and foster engagement. The topics covered are outlined in Table 1. Once the teaching strategies were implemented, their effectiveness was evaluated through formative assessments and qualitative feedback. They consist of two phases: wording and crossroads, which are described in Table 2.

Our primary investigation focused on students' definitions within the crossroad activity. Inspired by Michelini (2024) [43], we conducted qualitative analysis and reorganized the definitions into four conceptual categories. The first category, "Declarative Statements", consists of static, instructional statements that present terms in a straightforward manner. The second category, "Scientific Definitions", includes textbook-style definitions or those proposed by students that accurately reflect the scientific content associated with the term. The third category, "Recollection of (learning) Experiences", encompasses definitions based on prior knowledge or information learned during the activity. Lastly, the "Interpretative Statements" category includes definitions that reflect the students' personal interpretations of the physical meaning of the term. Each definition was assigned both a primary and secondary category, chosen from the four listed above, to capture the lexical and semantic nuances used by the students.

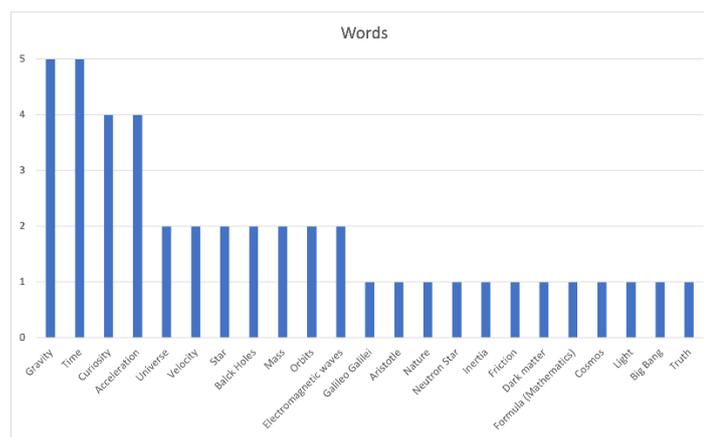

Fig. 1 Words listed by students and corresponding frequency (on the y-axes).

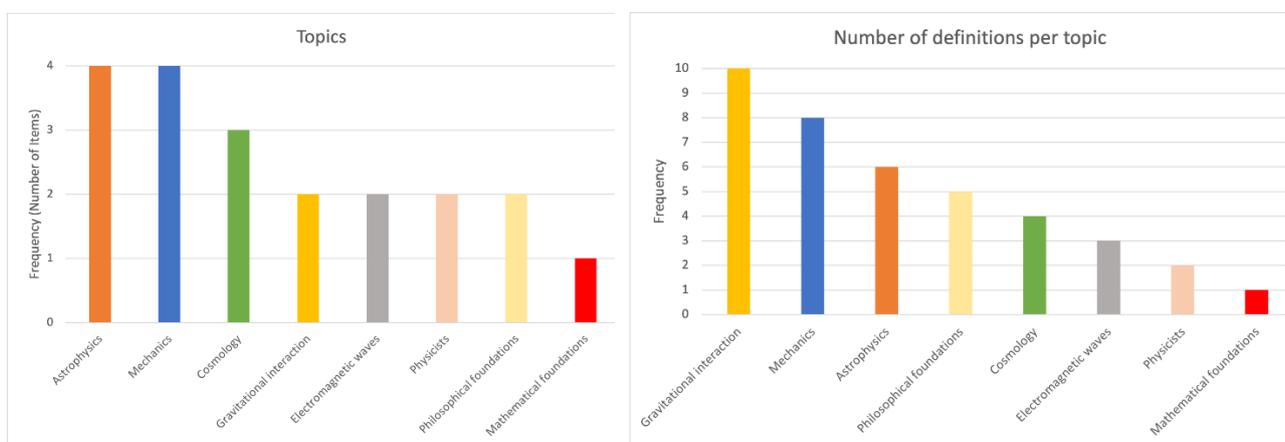

Fig. 2 Overall topics (on the left) and number of definitions per topic given by students (on the right).

Table 3. The distribution of definitions according to primary and secondary categories. Numbers represent the percentage of statements fitting each specific category. The highest percentages are in bold.

| GLOBAL PERCENTAGE | | Secondary category | | | |
|---|---|---|---|---|---|
| | | Declarative Statements | Scientific Definitions | Recollection of (Learning) Experiences | Interpretative statements |
| Primary category | **Declaritive statements** | / | 2.4 | 7.3 | **17.1** |
| | **(Scientific) Definitions** | 7.3 | / | **29.3** | 0 |
| | **Recollection of (learning) experiences** | 2.4 | 7.3 | / | 0 |
| | **Interpretative statements** | 2.4 | **19.5** | 4.9 | / |

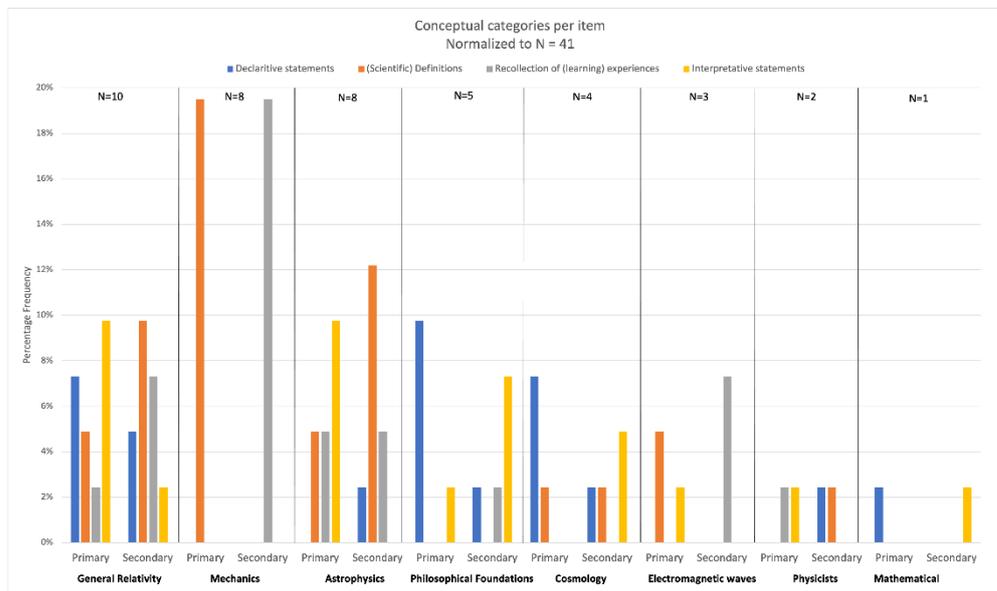

Fig. 3 Conceptual categories for each of the topics chosen by students. Each group of statements is normalized to the total number of definitions (n=41) collected.

| Gravità | Forza perpendicolare al piano |
|---|---|
| | Curvatura dello spazio correlato alla massa |
| | Ciò che permette ai corpi di non fluttuare. All'inizio si pensava che alla base di essa ci fosse la somiglianza dei corpi, il simile attrae il simile, ma grazie a Newton abbiamo scoperto che essa è applicata su ogni corpo e che i corpi non cadono in base all'altezza da cui cadono e varia la traiettoria. |
| | Fenomeno causato dalla curvatura del tessuto spazio-temporale a causa di gravi, o masse, molto grandi. Percepito ai nostri occhi come attrazione tra due corpi |
| | Forza attrattiva tra due masse |
| Tempo | Unità di misura relativa, dipendente allo spazio e alla gravità (se siamo vicini ad un grande centro di gravità, il "tempo" passa più lentamente). Escludendo tempo quantistico e tempo termico. |
| | Può essere considerato come una dimensione in cui gli eventi si susseguono ed è inoltre relativo poiché cambia in base alla velocità e alla gravità. |
| | Vicino a un buco nero scorre più lentamente |
| | Disegno: cono luce |
| | Scorre incessantemente, non può andare indietro (è relativo) |

| Accelerazione | Variazione di velocità di un determinato istante di tempo. La sua formula è (Delta v)/(delta t) |
| --- | --- |
| | Relationship between the speed and the time |
| | Rapporto tra lo spazio percorso e il tempo |
| | Rapporto tra velocità del corpo preso in considerazione ed intervallo di tempo |
| Velocità | Rapporto tra spazio percorso e tempo impiegato a percorrerlo |
| | Rapporto tra spazio percorso e tempo impiegato a percorrerlo |
| Inerzia | Resistenza di un corpo a ciò che viene causato da una forza |
| Attrito | Forza che si oppone al movimento del corpo |
| Stelle di neutroni | Una delle possibili conseguenze delle morti di una stella, che variano in base alla massa della nebulosa di partenza |
| Stella | Sono corpi celesti formati da gas caldo e plasma tutto compattato dalla gravità. Dentro il plasma ci sono particelle sparse; a differenza degli altri pianeti brillano di luce propria. |
| | Corpi celesti che emettono luce propria all'interno dei quali avvengono processi termonucleari. Sono fatte di gas e plasma |
| Buco nero | Distorsione dello spazio al centro della nostra galassia |
| | Corpo celeste con gravitazione molto alta dove non può uscire nulla (compresa la luce); è dovuto da esplosioni di più masse |
| Materia Oscura | Particelle non ancora scoperte e mai osservate |
| Curiosity | Forza invisibile che muove l'uomo verso l'ignoto e la verità |
| | Anima la ricerca |
| | Desiderio di conoscere |
| | Alimenta la mente umana |
| Natura | Tutta la materia che ci circonda (foreste, animali, ecc) |
| Universo | Spazio infinito in continua espansione |
| | Insieme dei corpi celesti che circonda la terra |
| Cosmo | Insieme di corpi celesti nell'universo e l'universo stesso |
| Big bang | Grande esplosione primordiale |
| Onde elettromagnetiche | Trasporto di energia con campi che si propagano in modo perpendicolare l'uno rispetto all'altro |
| | Noi vediamo una piccola fascia, ma fa parte di un grande insieme che noi non percepiamo (elettromagnetiche) |
| Luce | È un'onda che si propaga nel vuoto della quale noi vediamo solo uno spettro |
| Formula | Equazione che consente di calcolare un valore o misura |
| Galileo Galilei | Aveva un sogno: scrivere la matematica dell'universo. Il suo pensiero si basava sulle sensate esperienze. Riprende gli studi di un grande filosofo |
| Aristotele | Di chi è rimasta la concezione per 2000 anni |

Fig. 4. The distribution of collected definitions according to each identified macro-topic. Words are in Italian. Colors refer to the list in Fig. 2.

**Results**

Regarding the "wording" phase, the list of words chosen by the students is presented in Fig. 1. Figures 2 and 3 display the overall distribution of topics and the number of definitions associated with each topic. Figure 4 provides examples of students' definitions (in Italian). The results of the categorization of students' definitions are summarized in Table 3. Specifically, in 17.1% of cases, students who provided a definition in the form of a static statement based their response on a free interpretation of the term. In 29.3% of cases, students who proposed a scientific definition grounded their responses in knowledge acquired either during the activity (which focused on gravitational physics, astrophysics, and the philosophical aspects of physics) or from prior schooling in classical physics. An example of this was a student who defined acceleration as "the change in velocity over a given period of time." Moreover, in 19.5% of cases, students offering a free definition provided an interpretation that was consistent with the scientific definition of the term. For instance, one student defined time as "a relative unit of measure, dependent on space and gravity, where time passes more slowly near a gravitational center." Concerning the highest value in Table 3, a bias towards "Mechanics" was observed (see words and definitions in blue in Fig. 4); excluding this, the percentage dropped to 9.8%.

**Discussion**

The analysis of student definitions in physics education revealed valuable insights. The categorization model from [43] closely aligns with our findings, with two adjustments made: "quoted statements" was revised to "Scientific Definitions," and "Recollection of experiences" became "Recollection of learning experiences" to better suit our high school student sample.

When defining concepts like gravity and the Big Bang, students often relied on textbook or online sources, resulting in creative reconstructions of terms. Some referenced historical figures like Aristotle and Galileo, indicating that students drew heavily from prior knowledge. Misconceptions were common, particularly regarding gravity, with one group defining it as "what keeps bodies from floating," and another describing it as a force perpendicular to the plane. The Big Bang was often referred to as a "primordial big explosion," and students frequently used metaphors like "time is fluid" to describe abstract concepts. Tautological definitions and linguistic challenges also emerged, reflecting difficulties in articulating scientific ideas.

However, students in lower grades, particularly third-year students, provided definitions more aligned with contemporary physics, suggesting they were more open to modern scientific ideas. Some students even remarked, "But this is philosophy!" indicating a perceived overlap between physics and philosophy, which invites further reflection on the boundaries between the two disciplines from historical, epistemological, and educational perspectives [17,24,25].

The inquiry-based methodology of the teaching sequence fostered curiosity and guided students toward a structured understanding of scientific principles. By addressing common challenges in learning concepts like general relativity and modern cosmology, the historical approach enhanced students' understanding of gravitational physics while promoting critical thinking and appreciation for the scientific process. The use of debates, Q&A sessions, and formative assessments allowed for iterative refinement of teaching practices, ensuring meaningful student engagement and supporting conceptual change. This confirms previous results on the efficacy of such approach in teaching and learning of contemporary physics topics in high school [3,26].

**Conclusion**

This paper proposes a TLS designed to introduce high school students to contemporary physics topics. Based on the MER and DBL frameworks, our results show that the TLS effectively contextualizes the study of physics through its historical evolution, highlighting key controversies, challenges, and questions, while stimulating curiosity and cooperative learning (RQ1). Additionally, formative assessment tools, including inquiry, and game-based approaches, effectively track students' learning processes and can be easily applied in informal contexts (RQ2). The exploration of language and semiotic registers revealed patterns of reasoning, mental associations, and potential misconceptions (RQ3). The semiotic aspect was crucial. During the "wording" phase, students reflected on metaphors, analogies, and the limits of describing physical phenomena. Understanding why students select specific words helps them better articulate and discuss physics concepts, fostering deeper engagement.

Presenting contemporary physics topics from a historical and epistemological perspective proved effective in enhancing students' learning. Tracing key scientific theories helped students understand the milestones that shaped our contemporary knowledge of gravitational physics. This approach not only deepened their understanding of essential concepts but also encouraged critical thinking about the nature of science and its evolution. It also engaged students in the scientific method, prompting them to explore fundamental questions about the universe.

Formative evaluations offered important insights into students' learning processes and the influence of teaching methods on their comprehension. The feedback gathered from these evaluations informed ongoing modifications to TLS, showing its potentiality in facilitating conceptual change. Classroom discussions revealed that students found the activity engaging and enjoyable, with particular appreciation for its interdisciplinary nature. Many students expressed interest in similar future initiatives, recommending the integration of such approaches into the curriculum. Our findings suggest that the TLS and our interdisciplinary approach could be valuable tools for engaging students in STEM and guiding them toward academic careers. Future developments will include a quantitative study with a larger sample to more comprehensively monitor students' conceptions of contemporary physics. Additionally, a questionnaire will assess the activity's effectiveness, focusing on its impact on motivation and engagement.

**References**

[1] Deruelle N, Uzan J P, Relativity in Modern Physics, *Oxford University Press* (2018)
[2] Tuveri M, Steri A, The 'Universe in a Box': a hands-on activity to introduce primary school students to cosmology, *Phys. Educ.* 60 (2025) 015002
[3] Tuveri M., Steri A., and Fadda D., Using storytelling to foster the teaching and learning of gravitational waves physics at high-school, *Phys. Educ.* 59 (2024) 045031
[4] Kragh H. S., Conceptions of Cosmos – From Myths to the Accelerating Universe: A History of Cosmology, *Oxford University Press* (2007)
[5] Blum, R. Lalli, and J. Renn, The Reinvention of General Relativity: A Historiographical Framework for Assessing One Hundred Years of Curved Space-time, *Isis*, Vol. 106, No. 3 (2015), 598-620, The University of Chicago Press on behalf of The History of Science Society,
[6] H. Goenner, A golden age of general relativity? Some remarks on the history of general relativity, Gen Relativ Gravit (2017) 49:42, DOI 10.1007/s10714-017-2203-1
[7] Malcom S. Longair, A Brief History of Cosmology, Carnegie Observatories Astrophysics Series, Vol. 2: Measuring and Modeling the Universe, 2004 ed. W. L. Freedman (Cambridge: Cambridge Univ. Press)
[8] David W. Hughes and Richard de Grijs, The Top Ten Astronomical "Breakthroughs" of the 20th Century, Communicating Astronomy with the Public Journal, Cap V Vol. 1, No. 1, October (2007)
[9] S. Vitale, The first 5 years of gravitational-wave astrophysics, Science 372 (2021) 6546, abc7397,
[10] Abbott B P et al., Observation of gravitational waves from a binary black hole merger, *Phys. Rev. Lett.* 116 (2016) 061102
[11] Maggiore M. et al., Science case for the einstein telescope, *J. Cosmol. Astropart. Phys.* 03 (2020) 050
[12] Branchesi M. et al., Science with the einstein telescope: a comparison of different designs J. *Cosmol. Astropart. Phys.* 07 (2023) 068
[13] Punturo M et al., The einstein telescope: a third-generation gravitational wave observatory *Class. Quantum Grav.* 27 (2010) 194002
[14] M. Morganti. Filosofia della fisica. Un'introduzione, *Carocci Editore* (2016)
[15] J. L. Heilbron, The history of physics. A very short introduction, *Oxford University Press* (2018)
[16] Kragh H., Big bang: the etymology of a name, *Astron. Geophys.* 54 (2013) 28–30
[17] Kragh H., On modern cosmology and its place in science education, *Sc. and Ed.* 20 (2011) 343–57
[18] Corni, F., The role of metaphors in teacher education in physics., *Phys. Teach. Ed.,* (2023), 3-24.

[19] Kokkotas, P., et al., The role of language in understanding the physical concepts. *Nonlinear Analysis, Theory, Methods & Applications,* 30(4), (1997), 2113-2120.

[20] Kramar, N., et al., From intriguing to misleading: The ambivalent role of metaphor in modern astrophysical and cosmological terminology. *Amazonia Investiga*, 10(46), (2021), 92-100.

[21] Tiberghien, A. et al., Science and technology education at cross roads: Meeting the challenges of the 21st century. The second Conference of EDIFE and the Second IOSTE Symposium in Southern Europe. (2005)

[22] Duval, R. Semiosis et pensée humaine: sémiotiques registres et apprentissages intellectuels. Berna: Peter Lang. (1995)

[23] Brookes, D. T., Etkina, E. Using conceptual metaphor and functional grammar to explore how language used in physics affects student learning. *Phys. Rev. Phys Ed. Res.*, 3, (2012) 010105.

[24] Fouad, Khadija E., Heidi Masters, and Valarie L. Akerson. "Using history of science to teach nature of science to elementary students." *Science & Education* 24 (2015): 1103-1140.

[25] Gooday, Graeme, et al. "Does science education need the history of science?" *Isis* 99.2 (2008): 322-330.

[26] Tuveri M. et al., Promoting the learning of modern and contemporary physics in high schools in informal and non-formal contexts, *Nuovo Cimento C* 46 (2023) 6

[27] Gorghiu L. M. et al., The Role of Non-formal Activities on Familiarizing Students with Cutting-Edge Science Topics, in Proceedings of the Education, Reflection, Development, Fourth Edition (ERD 2016), 08- 09 July 2016, Babes-Bolyai University Cluj-Napoca- Romania (Future Academy) 2016, http://dx.doi.org/10.15405/epsbs.2016.12.56

[28] Eshach H., Bridging In-school and Out-of-school Learning: Formal, Non-Formal, and Informal Education., *J Sci Educ Technol* 16, (2007) 171–190 https://doi.org/10.1007/s10956-006-9027-1

[29] M. Komorek, and R. Duit (2004), The teaching experiment as apowerful method to develop and evaluate teaching and learning sequences in the domain of non-linear systems, *Int. J. of Sci. Ed.,* 26:5, 619-633, DOI: 10.1080/09500690310001614717

[30] P. Lijnse, Didactical structures as an outcome of research on teaching–learning sequences?, *Int. J. of Sci. Ed.,* 26:5, (2004), 537-554

[31] O. R.Battaglia, A. A. Gallitto , G. Termini, and C. Fazio, Outcomes of a Teaching Learning Sequence on Modelling Surface Phenomena in Liquid, *Educ. Sci.,* 13, (2023), 425.

[32] A. Tiberghien, J. Vince, and P. Gaidioz, Design-based research: case of a teaching sequence on mechanics. *Int. J. of Sci. Ed.,* Taylor & Francis (Rout- ledge), 31 (17), (2009), pp.2275-2314.

[33] C. Buty, A. Tiberghien, and J. F. Le Maréchal, Learning hypotheses and an associated tool to design and to analyse teaching–learning sequences, *Int. J. of Sci. Ed.,* 26(5), (2004), 579-604,

[34] A. Marzari, M. Di Mauro, T. Rosi, P. Onorato, M. Malgieri, Investigating the Principle of Relativity and the Principle of Equivalence in Classical Mechanics: Design and Evaluation of aTeaching–Learning Sequence Based on Experiments and Simulations. *Educ. Sci.*, 13 (2023) 712.

[35] A. Tiberghien, C. Buty, and J.-F. Le Maréchal, Physics teaching sequences and students' learning, Science and Technology Education at cross roads: meeting the challenges of the 21st century. The second Conference of EDIFE and the Second IOSTE Symposium in Southern Europe. (2005).

[36] M. Mèheut, and D. Psillos, Teaching–learning sequences: aims and tools for science education research, *Int. J. Sci. Educ.,* 26(5), 2004, 515–535

[37] Anderson, T., & Shattuck, J. (2012). *Design-Based Research: A Decade of Progress in Education Research?* Educational Researcher, 41(1), 16-25.

[38] Barab, S., & Squire, K. (2004). *Design-Based Research: Putting a Stake in the Ground.* The Journal of the Learning Sciences, 13(1), 1-14.

[39] Brown, A. L. *Design Experiments: Theoretical and Methodological Challenges in Creating Complex Interventions in Classroom Settings., J. of the Learn. Sci.,* 2(2), (1992)141-178.

[40] Psillos, D., & Kariotoglou, P., Teaching and Learning in the Science Laboratory, Kluwer Academic Publishers (Springer), Boston (2002)

[41] Leach, J., & Scott, P. (2002). Designing and Evaluating Science Teaching Sequences: An Approach Drawing upon the Concept of Learning Demand and a Social Constructivist Perspective on Learning., *Stud. in Sci. Ed.*, 38, 115-142.

[42] Tiberghien, A., Vince, J., & Gaidioz, P. Design-Based Research: Case of a Teaching Sequence on Mechanics. *Int. J. of Sci. Ed.*, 31(17), (2009) 2275-2314.

[43] Michelini M., Research-Based MEDS Model to Build Prospective Primary Teachers Professional Competence in Physics Education. In: Aydiner, E., Sidharth, B.G., Michelini, M., Corda, C. (eds) Frontiers of Fundamental Physics FFP16. FFP 2022. Springer Proceedings in Physics, vol 392. Springer, Cham., (2024)